\documentclass{sig-alternate-2013}
\usepackage{url}
\usepackage{amssymb}
\usepackage{enumitem}
\usepackage{verbatim}
\usepackage{pifont}

\usepackage{framed}
\usepackage{multirow}
\usepackage{array}
\usepackage{placeins}
\usepackage{balance}
\usepackage{rotating}
\usepackage{booktabs}
\usepackage{rotating}
\usepackage{hhline}
\usepackage{hyphsubst}
\usepackage{amsmath}
\graphicspath{{images/}}
\usepackage{times}
\usepackage{enumitem}
\usepackage{float}

\newfont{\mycrnotice}{ptmr8t at 7pt}
\newfont{\myconfname}{ptmri8t at 7pt}
\let\crnotice\mycrnotice%
\let\confname\myconfname%

\permission{Permission to make digital or hard copies of all or part of this work for personal or classroom use is granted without fee provided that copies are not made or distributed for profit or commercial advantage and that copies bear this notice and the full citation on the first page. Copyrights for components of this work owned by others than ACM must be honored. Abstracting with credit is permitted. To copy otherwise, or republish, to post on servers or to redistribute to lists, requires prior specific permission and/or a fee. Request permissions from permissions@acm.org.}
\conferenceinfo{Under Review.}{}

\begin{document}


\title{Twitter in Academic Conferences: Usage, Networking and Participation over Time}

\numberofauthors{4} 
%
\author{
Xidao Wen\\
       \affaddr{University of Pittsburgh}\\
       \affaddr{Pittsburgh, USA}\\
       \email{xiw55@pitt.edu}
\alignauthor
Yu-Ru Lin\\
       \affaddr{University of Pittsburgh}\\
       \affaddr{Pittsburgh, USA}\\
       \email{yurulin@pitt.edu}
\and
Christoph Trattner\\
       \affaddr{Know-Center}\\
        \affaddr{Graz, Austria}\\
       \email{ctrattner@know-center.at}
\alignauthor
Denis Parra\\
       \affaddr{PUC Chile}\\
       \affaddr{Santiago, Chile}\\
       \email{dparra@ing.puc.cl}
}

\maketitle
\begin{abstract}
Twitter is often referred to as a backchannel for conferences. While the main conference takes place in a physical setting, attendees and virtual attendees socialize, introduce new ideas or broadcast information by microblogging on Twitter. In this paper we analyze the scholars' Twitter use in 16 Computer Science conferences over a timespan of five years. Our primary finding is that over the years there are increasing differences with respect to conversation use and information use in Twitter. We studied the interaction network between users to understand whether assumptions about the structure of the conversations hold over time and between different types of interactions, such as retweets, replies, and mentions. While `people come and people go', we want to understand what keeps people stay with the conference on Twitter. By casting the problem to a classification task, we find different factors that contribute to the continuing participation of users to the online Twitter conference activity. These results have implications for research communities to implement strategies for continuous and active participation among members.

\end{abstract}



\section{INTRODUCTION} \label{sec:introduction}
Twitter, as one of the most popular microblogging services, has been raised as the backchannel of the academic conferences \cite{ross2011enabled}. There is a considerable amount of research into understanding the users' behavior on Twitter during academic events. Researchers look into why people tweet \cite{ebner2009people,ross2011enabled}, and what people tweet from a small number of conferences \cite{ebner2009introducing,icswm2009letierce2010using,weller2011citation}. 

In this work, we collect data about a larger set of academic conferences over five consecutive years, and provide in-depth analysis on this temporal dataset. We are particularly interested in the following research questions:

\begin{description}[itemsep=3pt,parsep=3pt,topsep=3pt, partopsep=0pt]
  \item[RQ1:] Do users use Twitter more for socializing with peers or for
information sharing during conferences? How has such use of Twitter
during conferences changed over the years?
\end{description}
\begin{description}[itemsep=3pt,parsep=3pt,topsep=3pt, partopsep=0pt]
\item[RQ2:] What are the structures of conversation and information sharing
networks in individual conferences? Have these network structures
changed over time?
\end{description}
   \begin{description}[itemsep=3pt,parsep=3pt,topsep=3pt, partopsep=0pt]
 \item[RQ3:] Do users participate on Twitter for the same conference over
consecutive years? To what extent can we predict users'
future conference participation?
\end{description}

To answer these questions, we crawled a dataset that consists of the tweets from 16 Computer Science conferences from 2009 to 2013. We examined Twitter in conferences by characterizing their use through retweets, replies, etc. We studied the structure of conversation and information-sharing by deriving two networks from the dataset, conversation and retweet network. Furthermore, to understand the factors that drive users' continuous participation, we propose a prediction framework with usage and network metrics.

As a result of our analyses, we found: (i) an increasing trend of informational usage (urls and retweets) compared to the stable pattern of conversational usage (replies and mentions) of conferences on Twitter over time, (ii) the conversation network is more fragmented than the information network, and the former becomes more fragmented over the time, and (iii) that the number of timeline tweets, users' centrality in information networks, and number of contacts in conversation networks are the most relevant features to predict users' continuous participation. These results summarize the online user participation of a real research community, which in turn helps to understand how it is perceived in online social networks and whether it is able to attract recurring attention over time.



The rest of the paper is structured as follows: The next section surveys Twitter used as a backchannel in conferences. Then, section 3 describes the dataset in this study and how we obtained it. Section 4 presents the experiment setup, followed by section 5 which provides the results. Section 6 summarizes our findings and concludes our paper with discussion of the future work. \\
\section{RELATED WORK}


Several authors have studied Twitter usage in events as diverse as politics \cite{politicslarsson2012studying,politicstumasjan2010predicting}, sports \cite{sportslanagan2011using,sportszubiaga2011classifying}, and natural disasters \cite{crisisabel2012twitcident,crisismendoza2010Twitter}. However, research that studies Twitter as a backchannel in academic conferences is closer to our work. Ebner et al. \cite{ebner2009introducing} studied tweets posted during the ED-MEDIA 2008 conference, and they argued that micro-blogging can enhance participation in the usual conference setting. Ebner et al.\cite{ebner2009introducing} eventually conducted a survey over 41 people who used Twitter during conferences, finding that people who actively participate in conference via Twitter are not only physical conference attendants, and that the reasons people have to participate are sharing resources, communicating with peers, and establishing online presence \cite{ebner2009people}. 

Considering another area, Letierce et al. \cite{icswm2009letierce2010using} studied Twitter usage by the semantic web community during the conference ISWC 2009. Eventually, they extended this research to analyze three conferences \cite{letierce2010understanding} and they found that analyzing Twitter activity during the conference helps to summarize the event (by categorizing hashhtags and URLs shared), and that the way people share information online is affected by the use of Twitter. In a different study, Ross et al. \cite{ross2011enabled} investigated the use of Twitter as a  backchannel within the community of digital humanists. By studying three conferences in 2009 they found that the micro-blogging activity during the conference is not a single conversation but rather multiple monologues with few dialogues between users and that the use of Twitter expands the participation of members of the research community. With respect to applications, Sopan et al. \cite{sopan2012monitoring} created a tool that provides real-time visualization and analysis for backchannel conversations in online conferences. 

In \cite{WenCscw2014}, we investigated the use of Twitter in three conferences in 2012 related to user modeling research communities. We classified Twitter users into groups and we found that the most senior members of the research community tend to communicate with other senior members, and newcomers (usually masters or first year PhD students) receive little attention from other groups, challenging  Reinhardt's assumption \cite{ebner2009people} about Twitter being an ideal tool to include newcomers in an established learning community.

Compared to previous research and to the best of our knowledge, this article is the first one studying a larger sample of conferences (16 in total) over a period of five years (2009-2013). This dataset allows us to generalize our results to Information and Computer Science and also to analyze trends of Twitter usage over time.

\section{Dataset}
Previous studies on analyzing the tweets during conferences examined a small number of conferences \cite{desai2012tweeting,icswm2009letierce2010using}. For each conference, they collected the tweets that contain the conference official hashtag in its text, for example, \textit{\#kidneywk11}, \textit{\#iswc2009}. They produced interesting results of how users employ Twitter during the conference, but their results are limited considering that they analyze at most three conferences. On the other hand, we are interested in studying trends of the usage and the structure over time, where we aimed to collect a dataset of tweets from a larger set of conferences over several years. Following the list of conferences in Computer Science listed in \textit{csconf.net}, we used the Topsy API\footnote{http://topsy.com} to crawl tweets by searching for the conference hashtag within a two-week time window (seven days before the conference until seven days after the conference ended).

\textbf{Conference dataset}. For this study, we focused on the conferences that had Twitter activity from 2009 to 2013. The crawling process took two weeks in December, 2013. We aggregated 109,076 tweets from 16 conferences over last 5 years. 

\textbf{User-Timeline dataset}. We acknowledge that users would also interact with other users without the conference hashtag, and therefore we additionally constructed the timeline dataset by crawling the timeline tweets of those users who participated in the conference during the same period. Table 1 shows the statistics of our dataset and Table \ref{table:conferenceDataset} (in Appendix) shows detailed information about each conference.

\textbf{Random dataset}. Any pattern observed would be less relevant unless we compare with a baseline, because the change might be a byproduct of Twitter usage trend overall. Hence, we show the conference tweets trend in comparison with a random sampled dataset. Ideally, we would like to obtain a random sample from the full tweets data collection in each year to compare them with the conference tweets. There is research on random sampling approaches based on Twitter API and the comparison of these approaches, but none of them attempted to sample from historical tweets, because Twitter API does not provide access to them. To overcome this issue, we again used Topsy API, for which it claims to have full access to all historical tweets. We have three random processes to make sure our sample is random and representative of the tweets. We randomly picked 2/3 of all the hours in each year, randomly picked two minutes from the each hour as our search time interval, and randomly picked the page number in the returned search result. The query aims to search for the tweets with any one of the alphabetical characters (from a to z). The crawling process took two days in December, 2013. As each query returned us 100 tweets, we were able to collect 5,784,649 tweets from 2009 to 2013.

\begin{table}[t!]
\small
\vspace{2mm}
  \centering
  \setlength{\tabcolsep}{2pt}
    \begin{tabular}{ll||lllll} 
    \specialrule{.2em}{.1em}{.1em}
    \multicolumn{2}{c||}{}& 2009  & 2010 & 2011 & 2012& 2013 \\ \hline

                     	&\#Unique Users & 1,114 & 2,970 & 3,022 & \bf{5,590} & 5,085 \\
			&\#Conference Tweets & 8,125 & 18,107 & 19,308 & \bf{34,424} & 27,549 \\
			&\#Timeline Tweets & 228,923 & 608,308 & 589,084 & \bf{1,025,259} & 939,760 \\   \specialrule{.2em}{.1em}{.1em}
  \end{tabular}
	\vspace*{-1mm}  
  \caption{Properties of the dataset collected in each year.}
	\vspace*{-1mm}  
  \label{table:dataset1}
	\vspace*{-1mm}    
\end{table}


\section{Methodology}
In this section we describe our experimental methodology, i.e., the metrics used, analyses and experiments conducted to answer the research questions:\\
 
\subsection{Analyzing the Usage}
We examined the use of Twitter during conferences by defining the measurements from two aspects: information usage and conversation usage. We want to use these measures to understand different usage dimensions and whether they have changed over time.

\textbf{Conversation usage}. With respect to using Twitter as a medium for conversations, we defined features based on two types of interactions between users: Reply and Mention ratios. For instance, @Alice can reply to @Bob, by typing `@Bob' at the beginning of a tweet, and this is recognized as a reply from @Alice. @Alice can also type @Bob in any part of her tweet except at the beginning, and this is regarded as a mention tweet. We computed the Reply Ratio to measure the proportion of tweets categorized as replies and the Mention Ratio respectively, but considering mentions.

\textbf{Information usage}. From the informational aspect of Twitter use during conferences, we computed two features to measure how it changed over the years: URL Ratio and Retweet Ratio. Intuitively, most of the URLs shared on Twitter during conference time are linked to additional materials such as presentation slides, publication links, etc. We calculated the URL Ratio of the conference to measure which proportion of tweets are aimed at introducing information to Twitter. The URL Ratio is simply the number of the tweets with `http:' over the total number of the tweets in the conference. 
The second ratio we used to measure informational aspects is Retweet Ratio, as the retweet plays an important role in disseminating the information within and outside the conference. We then calculated the Retweet Ratio to measure the proportion of tweets being shared in the conference. To identify the retweets, we followed a fairly common practice \cite{boyd2010tweetrt}, and used the following keywords in the queries: 'RT @', 'retweet @', 'retweeting @', 'MT @', 'rt @', 'thx @'.

Similarly, we computed the same features on the random dataset, as we wanted to understand if the observations in conference differ from the general usage on Twitter.

\begin{table*}[t!]
\small
\vspace{2mm}
  \centering
  \setlength{\tabcolsep}{3pt}
    \begin{tabular}{l|l||lllll} 
    \specialrule{.2em}{.1em}{.1em}
    \multicolumn{2}{c||}{Feature}& 2009  & 2010 & 2011 & 2012& 2013 \\ \specialrule{.1em}{.05em}{.05em}
    \multirow{7}{*}{\parbox{0.1cm}{\centering{\begin{sideways}\centering{Conversation}\end{sideways}}}}
&\#Nodes  &$ 165.313 \pm{50.358}$&$ 323.688 \pm{100.481}$&$ 385.625 \pm{100.294}$&$ 649.313 \pm{202.518}$&$ 622.188\pm{142.485}$ \\
&\#Edges  &$ 342.188 \pm{126.758}$&$ 660.625 \pm{240.704}$&$ 688.500 \pm{227.768}$&$ 1469.000 \pm{643.431}$&$ 1157.813\pm{344.484}$ \\
&In/Out degree &$ 1.446 \pm{0.153}$&$ 1.567 \pm{0.161}$&$ 1.502 \pm{0.086}$&$ 1.646 \pm{0.144}$&$ 1.618\pm{0.088}$ \\
&Density &$ 0.044 \pm{0.020}$&$ 0.010 \pm{0.002}$&$ 0.007 \pm{0.001}$&$ 0.005 \pm{0.001}$&$ 0.004 \pm{0.000}$ \\
&Clustering Coefficient &$ 0.066 \pm{0.015}$&$ 0.086 \pm{0.014}$&$ 0.074 \pm{0.008}$&$ 0.070 \pm{0.009}$&$ 0.078 \pm{0.006}$ \\
&Reciprocity &$ 0.172 \pm{0.034}$&$ 0.210 \pm{0.029}$&$ 0.237 \pm{0.023}$&$ 0.195 \pm{0.022}$&$ 0.203\pm{0.017}$ \\
&\#WCCs &$ 4.750 \pm{0.911}$&$ 13.438	\pm{3.243}$&$16.625 \pm{4.118}$&$	26.750 \pm{6.956}$&$ 29.188\pm{5.930} $\\

    \hline

    \multirow{7}{*}{\parbox{0.1cm}{\centering{\begin{sideways}\centering{Retweet}\end{sideways}}}}
 &\#Nodes & $87.063\pm{30.005}$ &  $355.500\pm{107.412}$ & $476.813\pm{117.641}$ & $720.875\pm{210.047}$ & $734.375\pm{153.998}$ \\
&\#Edges & $116.375\pm{46.124}$ & $722.125\pm{258.400}$ & $940.938\pm{277.384}$ & $1676.250\pm{693.462}$ & $1431.625\pm{351.722}$ \\
&In/Out degree & $0.981\pm{0.102}$ & $1.607\pm{0.129}$ & $1.653\pm{0.114}$ & $1.760\pm{0.143}$ & $1.728\pm{0.094}$ \\
&Density & $0.121\pm{0.048}$ & $0.009\pm{0.001}$ & $0.006\pm{0.001}$ & $0.004\pm{0.000}$ & $0.003\pm{0.000}$ \\
&Clustering Coefficient & $0.051\pm{0.016}$ & $0.078\pm{0.010}$ & $0.063\pm{0.008}$ & $0.048\pm{0.008}$ & $0.060\pm{0.006}$ \\
&Reciprocity & $0.053\pm{0.018}$ & $0.066\pm{0.010}$ & $0.054\pm{0.005}$ & $0.058\pm{0.008}$ &$ 0.070\pm{0.006}$ \\
&\#WCCs & $6.250\pm{1.627}$	&$6.500\pm{1.780}$	&$5.375\pm{1.341}$	&$6.625\pm{1.326}$&	$6.625\pm{1.998}$ \\
 \specialrule{.2em}{.1em}{.1em}  

  
  \end{tabular}
  \vspace{-2mm}
  \caption{Descriptive statistics (mean$\pm$ SE) of network metrics for the retweet and conversation networks over time. Each metric is an average over the individual conferences.}
  \vspace{-1mm}  
  \label{table:networkProperty}
\end{table*}


\subsection{Analyzing the Networks}
To answer RQ2, we conducted network analysis inspired by Lin et al. \cite{journals/corr/LinKML13}, who constructed networks from different types of communications: hashtags, mentions, replies, and retweets; and used their network properties to model communication patterns on Twitter. We followed their approach and focused on two networks derived from our dataset: conversation network, and retweet network. We defined them as follows:
\begin{itemize}[itemsep=0pt,parsep=0pt,topsep=0pt, partopsep=0pt]
 \item[$\bullet$]Conversation network: We built the user-user network of conversations for \textit{every conference each year}. This network models the conversational interactions(replies and mentions) between pairs of users. Nodes are the users in one conference and one edge between two users indicates they have at least one conversational interaction during the conference. 
\item[$\bullet$]Retweet network: We derived the user-user network of retweets for \textit{each conference each year}, in which a node represents one user and a directed link from one node to another means the source node has retweeted the targeted one.
\end{itemize}

The motivation for investigating the first two networks comes from the study of Ross et al. \cite{ross2011enabled}, who stated that: a) the conference activity on Twitter is constituted by multiple scattered dialogues rather than a single distributed conversation, and b) many users' intention is to jot down notes and establish an online presence, which might not be regarded as an online \emph{conversation}. This assumption is also held by Ebner et al. \cite{ebner2009people}. To assess if the validity of these assumptions holds over time, we conducted statistical tests over network features, including the number of nodes, the number of edges, density, diameter, the number of weakly connected components, and clustering coefficient of the network \cite{Wasserman1994}.

We constructed both conversation and retweet networks from the users' timeline tweets in addition to the conference tweets, as we guessed that many interactions might happen between the users without using the conference hashtag. Therefore, these two datasets combined would give us complete a more comprehensive dataset. Furthermore, we filtered out the users targeted by messages in both networks who are not in the corresponding conference dataset to make sure these two networks only capture the interaction activities between conference Twitter users. 



\subsection{Analyzing Continuous Participation}
To understand which users' factors drive their continuing participation in the conference on Twitter, we trained a binary classifier with some features induced from users' Twitter usage and their network metrics. From our own experience of attending the same conference again in real, we know that one reason is that we had valuable experience in the past -- quality research and social connections -- to be more specific. We expect that a similar effect exists with respect to the continuous participation on Twitter. Users' decision of whether coming back to the conference possibly depends on their experience with the conference via Twitter in the past -- valuable information and meaningful conversations. To capture both ends of a user's Twitter experience, we computed the usage measures, as described in Section 4.1, and user's network position \cite{journals/corr/LinKML13} in each of the networks: conversation network and retweet network, as discussed in Section 4.2. Measures for user's network position are calculated to represent the user's relative importance within the network, including degree, in-degree, out-degree, HIT hub score \cite{Kleinberg99authoritativesources}, HIT authority score \cite{Kleinberg99authoritativesources}, PageRank score \cite{Page99thepagerank}, eigenvector centrality score \cite{bonacich1972factoring}, closeness centrality \cite{Wasserman1994}, betweenness centrality \cite{Wasserman1994}, and clustering coefficient \cite{Wasserman1994}. 

\textbf{Dataset}. We identified 14,456 unique user-conference participations from 2009 to 2012 in our dataset. We then defined a continuing participation if one user shows up again in the same conference he or she participated in via Twitter last year. For example, @Alice posted a tweet with `\#cscw2010' during the CSCW conference in 2010, we counted it as one continuing participation if @Alice posted a tweet with `\#cscw2011' during the CSCW conference in 2011. By checking these users conference participation records via Twitter in the following year, we identified 2,749 continuing participations. Following the suggestion by Guha et al. \cite{Guha04propagationof}, we constructed a dataset with 2,749 positive continuing participations and 2,749 negative continuing participations (random sampling \cite{He:2009:LID:1591901.1592322}). 

\textbf{Features}. In the prediction dataset, each instance consisted of a set of features that describe the user's information \textit{in one conference in one year} from different aspects, and the responsive variable was a binary indicator of whether the user came back in the following year. We formally defined the key aspects of one user's features discussed above, in the following:
\begin{itemize}[itemsep=0pt,parsep=0pt,topsep=0pt, partopsep=0pt]
    \item[$\bullet$]Baseline: This set only includes the number of timeline tweets and the number of tweets with the conference hashtag, as the users' continuous participation might be correlated with their frequencies of writing tweets. We consider this as the primary information about a user and will be included in the rest of the feature sets.

\begin{figure}[t!]
  \centering
\includegraphics[scale = 0.28]{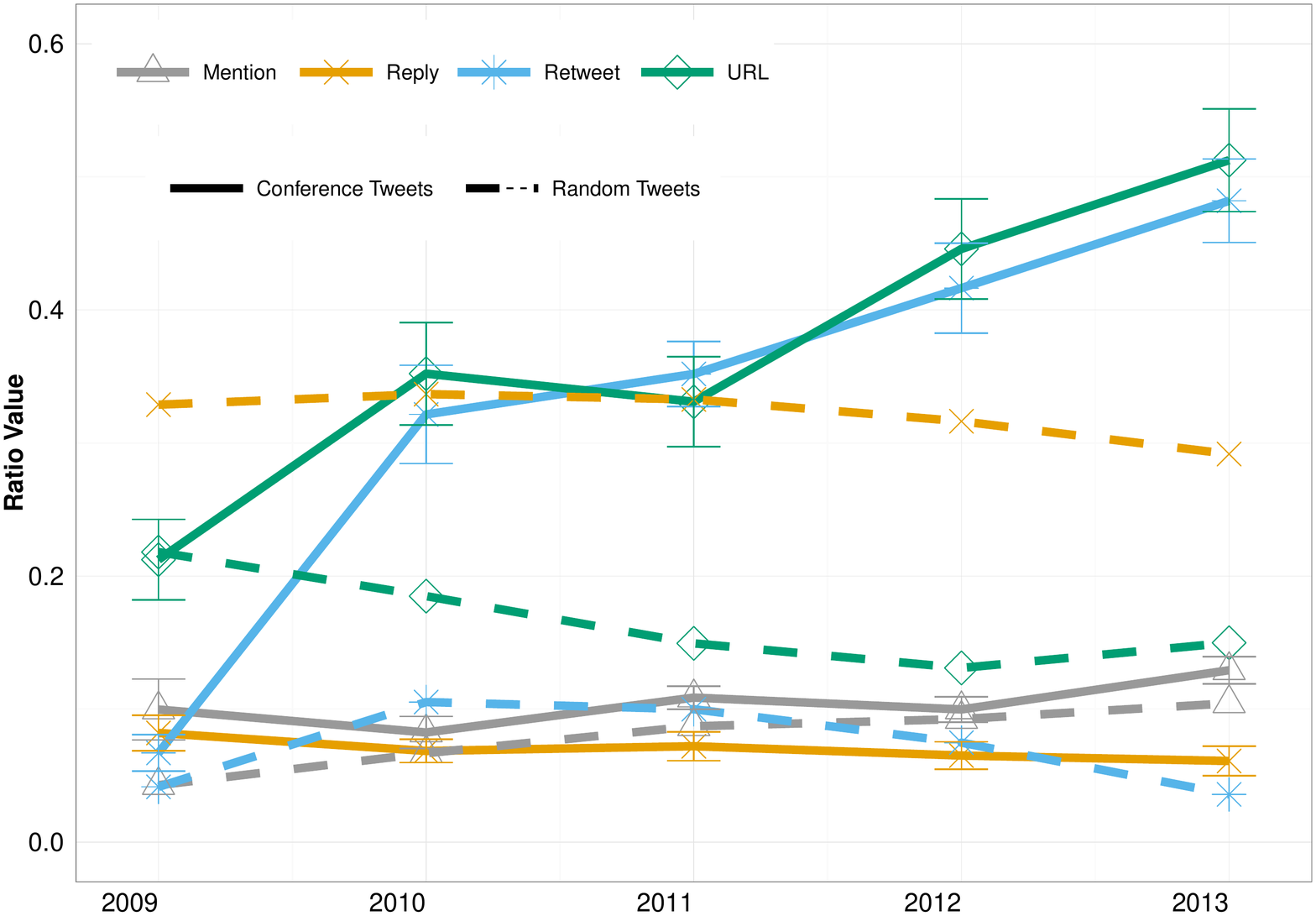}
	\vspace{-4mm}
  \caption[Usage pattern over the years]
   {Usage pattern over the years in terms of proportion of each category of tweets. Continuous lines represent conference tweets, dashed lines a random dataset from Twitter.}
	\label{fig:trendComparison}
	\vspace{-5mm}
\end{figure}
    
    \item[$\bullet$]Conversation: We built this set by including the conversation usage measures (Mention Ratio and Reply Ratio) and network position of the user in the conversation network.
    \item[$\bullet$]Information: Different from the previous one, this set captures the information oriented features, including the information usages (Retweet Ratio, URL Ratio) and user's position in the retweet network.
    \item[$\bullet$]Combined features: A set of features that utilizes all the features above to test the combination of them. 
\end{itemize}

\textbf{Evaluation}. To determine the importance of individual features, we used Information Gain as a measure for feature importance in WEKA \cite{Hall:2009:WDM:1656274.1656278}. Then we computed the normalized score of each variable's InfoGain value as its relative importance. To evaluate the binary classification model, we deployed different supervised learning algorithms and used the area under ROC curve (AUC) as our main evaluation metric to determine the performance of our feature sets. The evaluation was performed using 10-fold cross validation in the WEKA machine learning suite. 

\section{Results}
In the following sections we report on the results obtained for each of our analyses.
\subsection{Has usage changed?}

We answer this question by the results presented in Figure 1, which shows the overtime ratio values of different Twitter usage in the conferences accompanied by their corresponding values from the random dataset.

We can highlight two distinct patterns first. The trends we observed for the information usage ratios are similar. The Retweet Ratio increases (6.7\% in 2009, 48.2\% in 2013) over the years (one-way ANOVA, $p<.001$) along with URL Ratio (21.2\% in 2009, 51.3\% in 2013; one-way ANOVA, $p<.001$). Noticeably, Retweet Ratio rapidly increased from 2009 to 2010 but rather steadily gained afterward. We believe this could be explained by the Twitter interface being changed in 2009, when they officially moved `Retweet' button above the Twitter stream \cite{boyd2010tweetrt}. On the other hand, rather stable patterns can be observed in both conversational measures: Reply Ratio (8.2\% in 2009, 6.1\% in 2013) and Mention Ratio (10.0\% in 2009, 12.9\% in 2013). Therefore, as we expected, \textit{Twitter behaved more as an information sharing platform during the conference, while the conversational usage did not seem to change over the years}. 

Figure 1 also presents the differences between the ratio values in the conference dataset and the baseline dataset, as we want to understand if the trend observed above is the reflection of Twitter usage in general. During the conferences, we observed a higher Retweet Ratio and URL Ratio. We argue that it is rather expectable because of the nature of academic conferences: sharing knowledge and their valuable recent work. The Mention Ratio in the conference is slightly higher than it is in the baseline dataset, because the conference is rather an event of people interacting with each other in a short time period. However, we observe a significant difference in Reply Ratio. Users start the conversation on Twitter using the conference hashtag to some extent like all users do on Twitter, but most users who reply (more than 90\%) usually drop the hashtag. Although it is still a public discussion on Twitter, we guess that users tend to keep it semi-public to all their listeners, since these tweets are not visible to all the conference audience. However, a deeper analysis which is outside the context of this research should be conducted to assess this assumption, since in some cases users would drop the hashtag simply to have more characters available to reply with a larger message. 


\subsection{Has interaction changed?} 
Table 2 shows the evolution of the network measures. Each metric is an average over all the individual conferences in each year from 2009 to 2013. We first highlight the similar patterns over years observed from both networks: conversation network and retweet network. During the evolution, several types of network measures increase in both networks: (i) the average number of nodes; (ii) the average number of edges; and (iii) the average in/out degree. This suggests that more people are participating in the communication network.

Then, we compare these two networks in terms of the differences observed. Table 2 shows that the average number of weakly connected components in conversation network (\#WCC) grows steadily over time from 4.750 components in average in 2009 to a significantly larger 29.188 components in 2013, with the CHI conference being the most fragmented (\#WCC=87, \#Nodes=2117). However, the counterpart value in retweet network is almost invariable, staying between 5.375 an 6.625 in average. The former metric supports the assumption of Ross et al. \cite{ross2011enabled} in terms of the scattered characteristic of the activity network (i.e. multiple non-connected conversations). The \#WCC suggests that the retweet network is more connected than the conversation network.


Not surprisingly, the reciprocity in the conversation network is significantly higher than the one in the retweet network ($p<.001$ in all years; pair-wise t-test). This shows that the conversations are more two-way rounded interactions between pairs of users while people who get retweeted do not necessarily retweet back. Both results are rather expected. The mentions and replies are tweets sent to particular users, and therefore the addressed users are more likely to reply back due to social norms. Yet, the retweet network is more like a star network, and users in the center do not necessarily retweet back.


Moreover, we observe that the average clustering coefficient in conversation network is higher than the one in retweet network, in general. We think that two users who talked to the same people on Twitter during conferences are more likely to be talking to each other, while users who retweeted the same person do not necessarily retweeted each other. However, the significant difference is only found in 2012 ($p<.05$; t-test) and 2013 ($p<.001$; t-test). We tend to believe that it is the nature of the communication on Twitter, but we need more observations and further analysis to support this guess.

\subsection{What keeps the users returning?}

\begin{figure}[t!]
  \centering
\vspace{-2mm}  
\includegraphics[width=.45\textwidth]{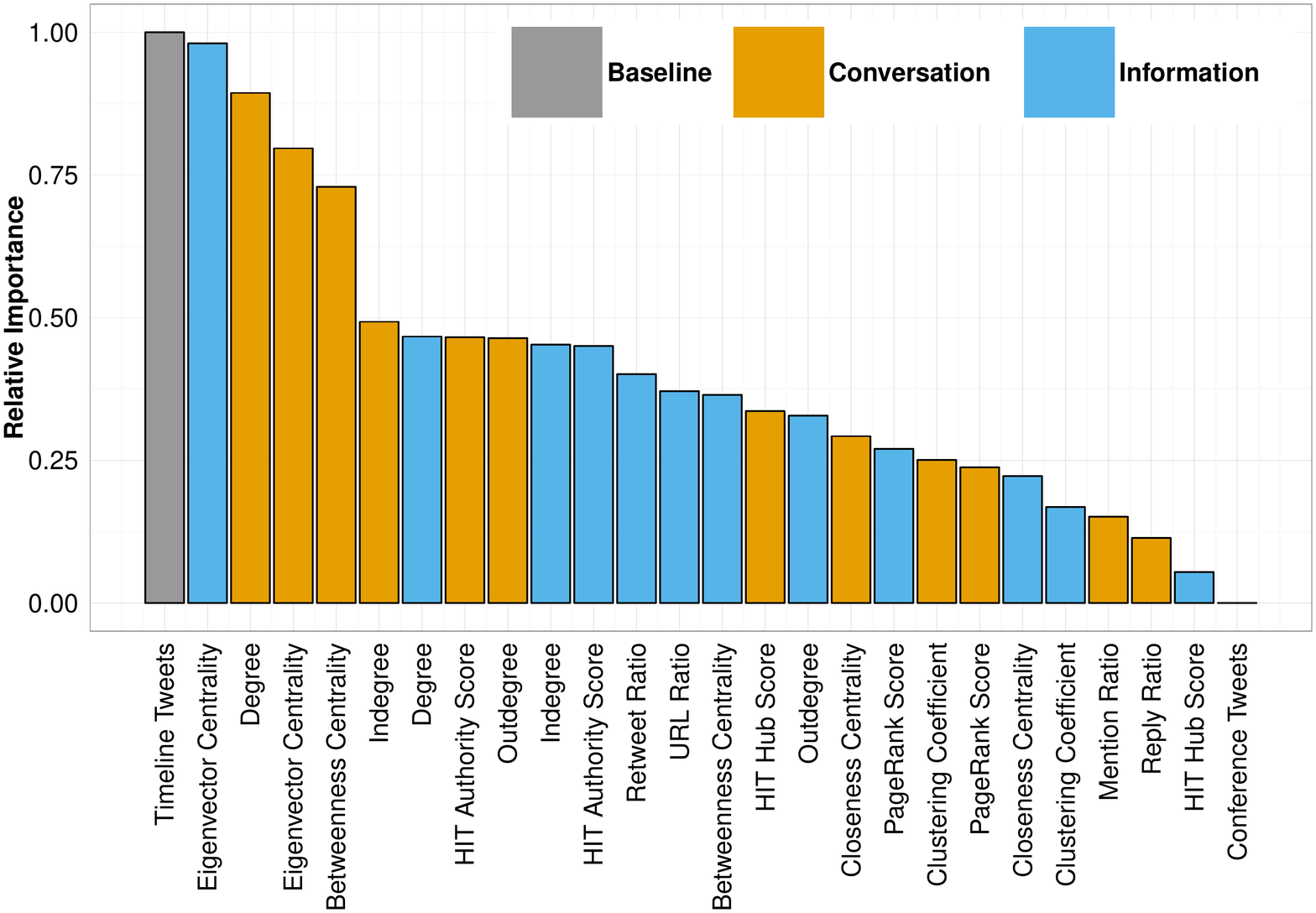}
\vspace{-4mm}
  \caption[Temporal trend comparison]
   {Relative Importance of the features.}
	\label{fig:InfoGain}
\end{figure}

The results of the prediction model are shown in Table 3, the Bagging approach achieved the highest performance across all the feature sets. Furthermore, baseline features achieved better performance when they are accompanied with conversation features than with the information features. This suggests that the conversation features have more predictive powers in inferring the users' continuous participation. Finally, the combination of all the features reached the best performance.

We further examine the importance of features in the combined set based on their information gain measures. Figure 2 shows the relative importance of different features. First, it is interesting two features in the baseline have distinct importances: the number of one user's conference tweets produced the least information gain, while the number of timeline tweets has the least information gain. This suggests that the user is more likely to return because she has been highly active on Twitter in general; the number of conference tweets tends to be an incident of many factors, for instance, users could have less conference tweets because she does not have time to tweet during the conference or the Internet connection is not available.  


We observe that the user's eigenvector centrality score from the information network has a higher importance (second lead in all the features). As the user with a higher eigenvector centrality score means that they and their neighbors have higher values in this measure, we interpret this as these users play important roles in information spreading during conferences. We conjecture these users are influential members in the conference so they are more likely to participate in the future. 

Furthermore, we observe that three measures of the conversation network are also prominent: degree centrality, eigenvector centrality, and betweenness centrality. While the first two indicate users' positions in the conversation network in terms of the number people they talked to and how important their neighbors are, the last one suggests that the user's role in bridging the conversation network. These suggest that users return because they have more diverse groups of friends on Twitter who are also active in conversing with others in the conference.

\begin{table}[t!]
\small
\vspace*{0mm}
  \begin{center}
\scalebox{0.9}{
  \setlength{\tabcolsep}{6pt}
    \begin{tabular}{l|l|lllll} 
    \specialrule{.2em}{.1em}{.1em}
    \multicolumn{2}{l||}{Feature Sets}& RF  & ADA & Bagging & LR & SVM \\ \hline
    \multicolumn{2}{l||}{Baseline}  & $0.654$      & $0.713$ & $\bf{0.716}$ & $0.714$ & $0.63$\\\hline
      \multicolumn{2}{l||}{Baseline+Conversation}  & $0.728$      & $0.754$ & $\bf{0.774}$ & $0.742$ & $0.659$      \\ \hline  
    \multicolumn{2}{l||}{Baseline+Information}  & $0.74$      & $0.761$ & $\bf{0.771}$ & $0.74$ & $0.655$      \\ \specialrule{.2em}{.1em}{.1em}  

    \multicolumn{2}{l||}{Combined}  & $0.753$      & $0.775$ & $\bf{0.787}$ &$0.754$     & $0.662$ \\

  \end{tabular}
  }
  \end{center}
   \vspace*{-2mm}
  \caption{Area under the ROC curve (AUC) for predicting continuing participation in the coming year with different feature sets and learning algorithms. The best algorithm for each feature set is highlighted. Methods used as Random forest (RF), Adaptive Boosting (ADA), Bagging, logistic regression (LR), and support vector machines (SVM).}
  \label{tab:overall_evaluation}
 \vspace*{-4mm}  
\end{table}

\section{Discussion and Conclusions}

In this paper, we investigated how Twitter has been used in academic conferences and how this has changed over the years. We attempted to study the usage, the network structures, and participation of the users during conferences on Twitter. We addressed our research with three questions.

To answer the first research question RQ1, we computed four features to quantitatively measure two aspects of Twitter use at conferences: information usage and conversation usage. Our results show that researchers are using Twitter as 2013 in a different way as they did in 2009 with respect to favoring more the information sharing value of Twitter, however their Twitter conversation interactions in 2013 do not differ much from five years ago.

Then, to answer the second research question RQ2, we constructed the conversation network and the retweet network for each conference and use network properties to measure how people interacted over the years. Our results show that with more people participating over time, the conversations turn scattered into small groups, while the information flow (retweets) stays mainly within a giant component.


Finally, to answer the third research question RQ3, we trained a binary classifier with features extracted from the Twitter usage, network positions of the user and achieve the best prediction performance using the combination of conversation features, information features, and baseline features. We also found that the most influential factors that drive users' continuous participation are being active on Twitter, being central in the information network and talking with more people. 


We acknowledge the limitation of this work. We conducted our analysis on sixteen conferences in the field of Computer Science, but the findings might not be generalized to all the computer science conferences since users in different sub-domains of this field might not have the same Twitter usage. In future work, we plan to extend our study to a larger set of conferences across different domains in the field such as Information Systems and Data Mining, in order to see whether users in these conferences behave differently in Twitter. 

\textbf{Acknowledgments:} This work is supported by the Know-Center and the EU funded project Learning Layers (Grant Nr. 318209).

\balance
\bibliographystyle{ieeetr}
\bibliography{ht14Twitter}




\appendix
\section{Complete dataset statistics}
\begin{table*}[ht!]
\small
 \vspace{4mm}
  \centering
  \setlength{\tabcolsep}{2pt}
    \begin{tabular}{lllllllllllllllllll}
    \specialrule{.2em}{.1em}{.1em}
Year & Measure & CHI & CIKM & ECTEL & HT & IKNOW & ISWC & IUI & KDD & RECSYS & SIGIR & SIGMOD & UBICOMP & UIST & VLDB & WISE & WWW \\ \hline
2009 & \# Unique Users & 299 & 7 & 55 & 42 & 37 & 255 & 1 & 11 & 52 & 91 & 1 & 20 & 38 & 15 & 5 & 311 \\
 & \# Tweets & 1860 & 36 & 487 & 247 & 173 & 1350 & 9 & 56 & 488 & 565 & 5 & 102 & 92 & 65 & 17 & 2581 \\
 & \# Retweets & 69 & 1 & 57 & 19 & 12 & 233 & 0 & 2 & 84 & 57 & 0 & 11 & 4 & 3 & 0 & 170 \\
 & \# Replies & 249 & 1 & 78 & 16 & 18 & 184 & 0 & 6 & 31 & 80 & 0 & 12 & 6 & 7 & 0 & 215 \\
 & \# Mentions & 160 & 2 & 32 & 22 & 18 & 203 & 0 & 4 & 48 & 64 & 0 & 10 & 9 & 6 & 7 & 158 \\
 & \# URL Tweets & 205 & 1 & 93 & 90 & 52 & 458 & 1 & 16 & 154 & 144 & 0 & 42 & 22 & 9 & 2 & 489 \\
  & \# Timeline Tweets & 37664 & 146 & 3132 & 3571 & 2577 & 37875 & 132 & 723 & 6179 & 31223 & 35 & 1226 & 3976 & 1149 & 263 & 100381 \\
  \hline
2010 & \# Unique Users & 787 & 53 & 32 & 43 & 50 & 179 & 46 & 47 & 138 & 84 & 37 & 89 & 53 & 20 & 562 & 1114 \\
 & \# Tweets & 5263 & 218 & 99 & 261 & 138 & 783 & 191 & 64 & 984 & 270 & 41 & 450 & 126 & 39 & 3006 & 5484 \\
 & \# Retweets & 1045 & 74 & 21 & 45 & 44 & 273 & 33 & 27 & 367 & 87 & 32 & 133 & 27 & 9 & 1280 & 1736 \\
 & \# Replies & 412 & 9 & 4 & 16 & 12 & 34 & 33 & 2 & 83 & 15 & 1 & 27 & 10 & 3 & 284 & 361 \\
 & \# Mentions & 467 & 21 & 4 & 26 & 20 & 58 & 17 & 6 & 88 & 17 & 0 & 25 & 16 & 0 & 207 & 1034 \\
 & \# URL Tweets & 810 & 50 & 31 & 92 & 54 & 378 & 50 & 21 & 285 & 84 & 32 & 143 & 50 & 22 & 527 & 1566 \\ 
   & \# Timeline Tweets & 199838 & 3243 & 1197 & 6999 & 1787 & 20359 & 5603 & 7157 & 9622 & 3459 & 2677 & 6480 & 6933 & 1797 & 107607 & 231618 \\
 \hline
2011 & \# Unique Users & 1207 & 153 & 58 & 119 & 97 & 452 & 61 & 175 & 194 & 83 & 63 & 74 & 70 & 102 & 37 & 604 \\
 & \# Tweets & 6042 & 898 & 314 & 457 & 372 & 3506 & 175 & 503 & 1517 & 176 & 187 & 363 & 179 & 507 & 98 & 2446 \\
 & \# Retweets & 1662 & 336 & 63 & 92 & 119 & 1619 & 64 & 204 & 669 & 93 & 49 & 121 & 51 & 147 & 47 & 995 \\
 & \# Replies & 627 & 37 & 33 & 74 & 24 & 211 & 8 & 35 & 144 & 4 & 6 & 52 & 19 & 10 & 2 & 152 \\
 & \# Mentions & 678 & 63 & 46 & 38 & 43 & 487 & 25 & 68 & 174 & 10 & 11 & 63 & 21 & 33 & 11 & 233 \\
 & \# URL Tweets & 1356 & 188 & 68 & 159 & 131 & 1478 & 59 & 209 & 481 & 59 & 47 & 125 & 50 & 94 & 75 & 718 \\ 
   & \# Timeline Tweets & 274026 & 13796 & 6161 & 14336 & 3926 & 62875 & 5006 & 20608 & 16888 & 5965 & 5337 & 16721 & 3333 & 8526 & 5132 & 138225 \\
 \hline
2012 & \# Unique Users & 1195 & 194 & 113 & 118 & 92 & 623 & 40 & 74 & 324 & 279 & 105 & 89 & 80 & 115 & 239 & 2653 \\
 & \# Tweets & 7590 & 747 & 544 & 381 & 638 & 2751 & 137 & 187 & 2256 & 1240 & 372 & 214 & 192 & 265 & 433 & 12150 \\
 & \# Retweets & 1981 & 352 & 197 & 190 & 92 & 1263 & 32 & 88 & 1270 & 660 & 99 & 72 & 100 & 118 & 275 & 5630 \\
 & \# Replies & 831 & 39 & 30 & 36 & 12 & 138 & 8 & 13 & 80 & 140 & 66 & 9 & 8 & 3 & 27 & 609 \\
 & \# Mentions & 732 & 58 & 94 & 54 & 55 & 322 & 10 & 10 & 305 & 68 & 31 & 29 & 11 & 13 & 54 & 1670 \\
 & \# URL Tweets & 1871 & 194 & 240 & 198 & 242 & 1053 & 51 & 116 & 971 & 416 & 125 & 88 & 110 & 144 & 369 & 5229 \\ 
   & \# Timeline Tweets & 135093 & 18005 & 45888 & 10724 & 8310 & 98034 & 4552 & 20259 & 36908 & 21008 & 9506 & 4547 & 9123 & 9146 & 48935 & 549884 \\
 \hline
2013 & \# Unique Users & 1526 & 785 & 123 & 59 & 98 & 424 & 67 & 447 & 240 & 373 & 88 & 141 & 225 & 103 & 207 & 1020 \\
 & \# Tweets & 7424 & 1525 & 707 & 121 & 271 & 2169 & 288 & 1324 & 1362 & 2059 & 238 & 591 & 533 & 241 & 479 & 4885 \\
 & \# Retweets & 2311 & 1095 & 290 & 35 & 114 & 1082 & 81 & 725 & 820 & 933 & 105 & 299 & 358 & 118 & 269 & 2492 \\
 & \# Replies & 883 & 27 & 53 & 10 & 21 & 96 & 22 & 40 & 62 & 392 & 16 & 18 & 10 & 9 & 13 & 179 \\
 & \# Mentions & 836 & 149 & 100 & 27 & 52 & 295 & 36 & 219 & 183 & 201 & 12 & 56 & 57 & 37 & 53 & 614 \\
 & \# URL Tweets & 2648 & 966 & 369 & 47 & 136 & 1190 & 101 & 550 & 797 & 721 & 80 & 361 & 386 & 120 & 430 & 2353 \\
   & \# Timeline Tweets & 249404 & 142472 & 12372 & 3542 & 13223 & 42405 & 4677 & 93997 & 25451 & 24609 & 29446 & 21478 & 19895 & 9798 & 62050 & 227908 \\
		\specialrule{.2em}{.1em}{.1em}		
    \end{tabular}
		 \caption{Detailed description of the dataset for each conference in each year.}
  \label{table:conferenceDataset}
\end{table*}

\end{document}